\documentclass[%
 reprint,
showpacs,
 amsmath,amssymb,
 aps,
pre,
]{revtex4-2}
 \usepackage{color}
 \usepackage{multirow}
 \usepackage{array}
\usepackage{graphicx}
\usepackage{dcolumn}
\usepackage{bm}

\begin{document}

\title[aaa]{Forecasting the El~Ni\~no type well before the spring predictability barrier}

\author{Josef Ludescher$^1$, Armin Bunde$^2$, Hans Joachim Schellnhuber$^1$}
\affiliation{
$^1$Potsdam Institute for Climate Impact Research (PIK), 14412 Potsdam, Germany
$^2$Institut f\"ur Theoretische Physik, Justus-Liebig-Universit\"at Giessen, D-35392 Giessen, Germany
}

\maketitle

{\bf
The El~Ni\~no Southern Oscillation (ENSO) is the most important driver of interannual global climate variability
and can trigger extreme weather events and disasters in various parts of the globe.
Depending on the region of maximal warming, El~Ni\~no events can be partitioned into 2 types, Eastern Pacific (EP) and Central Pacific (CP) events. The type of an El~Ni\~no has a major influence on its impact and can even lead to either dry or wet conditions in the same areas on the globe.
Here we show that the zonal difference $\Delta T_{WP-CP}$ between the sea surface temperature anomalies (SSTA) in the equatorial western Pacific and central Pacific gives an early indication of the type of an upcoming El~Ni\~no: When at the end of a year,  $\Delta T_{WP-CP}$ is positive, an event in the following year will be  probably an EP event, otherwise a CP event. Between 1950 and present, 3/4 of the EP forecasts and all CP  forecasts are correct.
When combining this approach  with a previously introduced climate-network approach, we obtain reliable forecasts for both the onset and the type of an event: at a lead time of about one year,  2/3 of the EP forecasts and all CP forecasts in the regarded period are correct. The combined model has considerably more predictive power than the current operational type forecasts with a mean lead time of about 1 month and should allow early mitigation measures.}

\bigskip
\bigskip
{\bf Significance statement:}
El~Ni\~no events represent anomalous episodic warmings, which can peak in the equatorial Central Pacific (CP events) or Eastern Pacific (EP events). Both types of events can have adverse and devastating impacts around the globe.
Currently, robust operational type forecasts have a mean lead time of about 1 month.
Here we show that the difference of the sea surface temperature anomalies between the equatorial western and central Pacific in December enables an early forecast of the type of an upcoming El~Ni\~no. Combined with a previously introduced climate network-based approach, both the onset and type of an upcoming El~Ni\~no can be efficiently forecasted. The lead time is about 1y and should allow early mitigation measures.
\bigskip
\bigskip

El~Ni\~no events are part of the El~Ni\~no-Southern Oscillation (ENSO) \cite{Clarke08, Sarachik10, Dijkstra2005, Wang2017,Timmermann2018, McPhaden2020}, which can be perceived as a self-organized quasi-periodic pattern in the tropical Pacific ocean-atmosphere system, featured by rather irregular warm (``El~Ni\~no'') and cold (``La Ni\~na'') excursions from the long-term mean state. Depending on the location of the peak warming of the SSTA, one usually distinguishes between Eastern Pacific (EP) and Central Pacific (CP) El~Ni\~no events.
The EP events exhibit their largest SSTA warming in the  eastern equatorial Pacific, while the CP events exhibit their largest warming westwards in the central equatorial Pacific.
The shift of the location of the maximum SSTA during CP El~Ni\~nos, compared to EP El~Ni\~nos, towards the central Pacific drives substantial shifts in atmospheric convection and circulation responses, which alter the location and intensity of temperature and precipitation impacts associated with El~Ni\~no around the globe  \cite{Larkin2005, Ashok2007,Weng2007, Wang2007, Taschetto2009, Frauen2014, Capotondi2015a, Freund2019, Wiedermann2021}.

Large EP events typically lead to strongly increased precipitation along the coast of Ecuador and Northern Peru, resulting in massive floodings and landslides, while CP events only lead to dry conditions in these already dry areas and also to dryer conditions in the Peruvian Andes (see, e.g., \cite{Lagos2008,Bazo2013}).  In India, particularly, CP events may lead to monsoon failures and thus to major droughts \cite{Kumar2006}. For more extensive discussions of the impacts of both  El~Ni\~no types, we refer to, e.g., \cite{Ashok2007, Weng2007, Wiedermann2021}.

These examples  demonstrate  that for  mitigating the societal impact of an El~Ni\~no event  by more targeted mitigation measures,  it is crucial to have  early operational forecasts not only for the event itself but also for the type of the event. Currently, this is not the case. For instance, Hendon et al. \cite{Hendon2009} found that the coupled ocean-atmosphere seasonal forecast model of the Australian Bureau of Meteorology is limited to less than 1 season in predicting the SSTA pattern of EP and CP events. In a more recent and comprehensive study, Ren et al. \cite{Ren2019} analyzed 6 operational climate models and found that only 2-3 can distinguish at 1 month lead time between CP and EP events. Zhang et al. \cite{Zhang2020} focused on hindcasts of the Climate Forecast System version 2 (CFSv2).
They found that the skill of the CFSv2 model was comparable to the one reported by Ren et al. \cite{Ren2019} and concluded that the CFSv2 model was broadly representative of the state-of-the-art forecast systems.
   
For comparison,  without distinction between CP and EP events, the current forecasts in operation (which are limited by the spring barrier) have a maximum lead time of about 6 months \cite{Barnston2012, McPhaden2020}. Forecasts based on a climate network \cite{Tsonis2006, Fan2020, Ludescher2021} approach have a lead time of about 1 year (see also Data and Methods Section) \cite{Ludescher2013,Ludescher2014}.
 
In the present study, we consider the period between 1950 and present, where reliable data on the El~Ni\~no events exist. We show that from the available SSTA data in the tropical Pacific, a precursor for CP and EP events can be obtained about 1 year before the peak of the event with high prediction skill.
When combined either with the conventional forecasts in operation or the recently established climate network approach, one arrives at significantly  better forecasts for the type of an El~Ni\~no event, with the same lead time as the conventional or the network approaches, respectively.
 
\subsection*{Classification of the El~Ni\~no types}
To identify the types of the 23 El~Ni\~no events between 1950 and present (here, the El~Ni\~no events  from 1986-1988 and 2014-2016 are counted as 1 event each), we have used 11 classification approaches
\cite{Larkin2005, Ashok2007, KaoYu2009, Kim2009, Kug2009, Yeh2009, RenJin2011, Takahashi2011, YuKim2013, Wiedermann2016, Feng2019}.
Most of these approaches rely on compressing/projecting the diverse spatial SSTA patterns in the tropical Pacific into simple scalar numbers, i.e., indices. The choice of which geographical areas, times of the year, or physical quantities to regard and
how to analyze the data leads to a large set of possible indices. Figure 1 summarizes the classification according to the 11 methods. For details, see the SI.
For the majority of El~Ni\~no events, in particular the strong ones, there is a high consensus about the type irrespective of the approach. 
For the events in 1969/70 and  1986-1988, there is lacking consensus, so we keep them as unidentified. For more details, see the SI.

\bigskip

\begin{figure}[]
\begin{center}
\includegraphics[width=\columnwidth]{./fig1.pdf}
\caption{Classification of the type of El~Ni\~no events.
The table summarizes the classification according to 11 approaches.
The numbers in the first row indicate the last two digits of the El~Ni\~no onset years. 
Eastern Pacific El~Ni\~nos are marked 'E' and shown in orange. Central Pacific El~Ni\~nos are marked 'C' and shown in blue. El~Ni\~no events  marked by 'M' are not  identified as pure EP or CP events, see SI.}
\label{fig1}
\end{center}
\end{figure}

\subsection*{Precursor for the type of the next El~Ni\~no event}
Due to the mostly easterly winds of the Walker circulation,  the climatological background state of the equatorial Pacific is characterized by a  temperature gradient from the cold tongue in the east to the warm pool in the west.
Accordingly, the  simplest conceptual models of ENSO,  for instance, the recharge-oscillator model  developed by Jin \cite{Jin1997}, divide  the Pacific into two regions, the western Pacific and the eastern Pacific. The models are able to describe the occurrence of  the canonical EP events but not of CP  events \cite{Fang2018}.

Since CP El~Ni\~nos are located in the central Pacific, it appears natural to extend the two-region conceptional models by including the central Pacific as a third region.
This region is important for the development of CP events since the zonal advective feedback is most effective here \cite{Capotondi2013, Fang2018}. Such three-region  models were developed by Fang and Mu \cite{Fang2018} and Chen et al. \cite{Chen2022}  and indeed allow for the occurrence of both EP and CP events \cite{Chen2022}.
Within their model, Chen et al. \cite{Chen2022} found that on decadal scales the frequency of CP events depended  particularly on the SST gradient between the western and central Pacific (which is directly related to the zonal advective feedback and highly correlated with the strength of the Walker circulation). This suggests that the occurrence of CP events may be related to the zonal SST difference between the western and central Pacific.

Here we follow this idea and study the temporal fluctuations of the zonal temperature difference between the western Pacific and central Pacific. We want to find out whether there are precursor phenomena that can be used to forecast the type of the next El~Ni\~no  well before its onset.

To be specific, we consider the monthly zonal SSTA difference $\Delta T_{WP-CP}$ between two equal-sized areas
in the west Pacific (120E-165E, 5N-5S) and the central Pacific (165E-150W, 5N-5S) (see Fig. 2). The width of both areas is identical to the widths of the Ni\~no3, Ni\~no4 and Ni\~no3.4 areas. The CP area nearly coincides with the Ni\~no4 area.
To account for the different effects of climate change in both regions, we use a trailing 30 y climatology, i.e., we calculate the temperature anomalies based only on the past of the considered point in time.

\begin{figure}[]
\begin{center}
\includegraphics[width=8.4cm]{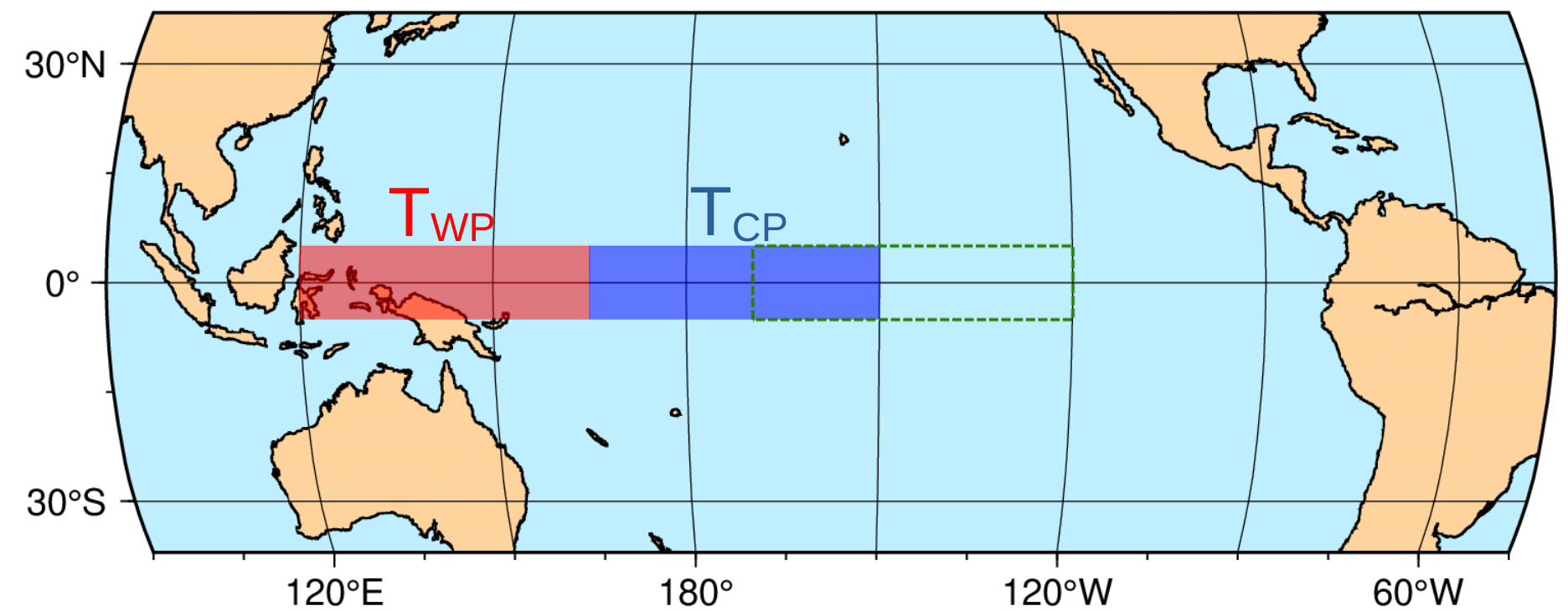}
\caption{The areas of the regarded sea surface temperature anomalies (SSTA).  The sign of the zonal difference $\Delta T_{WP-CP}$  between the SSTA  in the western equatorial Pacific and the SSTA in the central Pacific is predictive of the type of an upcoming El~Ni\~no event.
The red area represents the western warm pool, the blue area nearly coincides with the Ni\~no4 area, where CP El~Ni\~no events are centered. The dashed green rectangle shows the Ni\~no3.4 area.}
\label{fig2}
\end{center}
\end{figure}

\bigskip

\begin{figure*}[]
\begin{center}
\includegraphics[width=14cm]{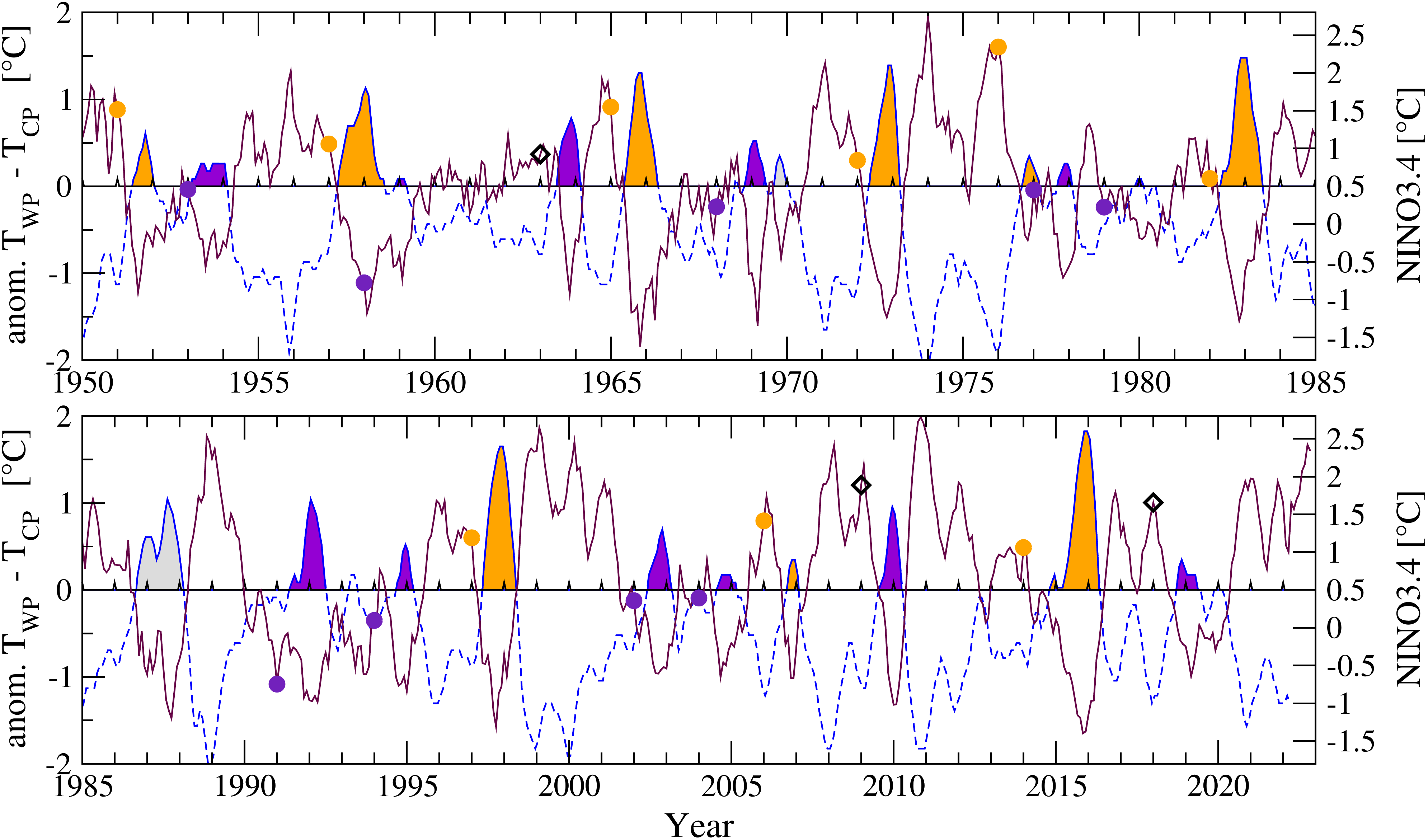}
\caption{SSTA based forecasting scheme for the  type of an El~Ni\~no event.
The figure shows the anomaly of the monthly sea surface temperature difference between the equatorial western Pacific (120E-165E, 5N-5S) and central Pacific (165E-150W, 5N-5S) (left scale, maroon line) and the ONI (right scale, blue dashed line). The shaded areas highlight El~Ni\~no episodes, in orange EP El~Ni\~nos and in violet CP El~Ni\~nos. The classification methods lead to a tie for the 1969/70 El~Ni\~no and the 1986-1988 multiyear El~Ni\~no (both in gray).  Positive values of  $\Delta T_{WP-CP}(t)$ at the end of a calendar year before an El~Ni\~no onset serve as precursor for an EP event, negative values as  precursor for a CP event.
Of the 21 type forecasts, 18 are correct: there are 9 correct predictions for an EP event (orange circles), 3 false predictions where instead of an EP event, a CP event occurred (black diamonds), and 9 correct predictions of a CP event (violet circles).}
\label{fig3}
\end{center}
\end{figure*}

\begin{figure*}[]
\begin{center}
\includegraphics[width=14cm]{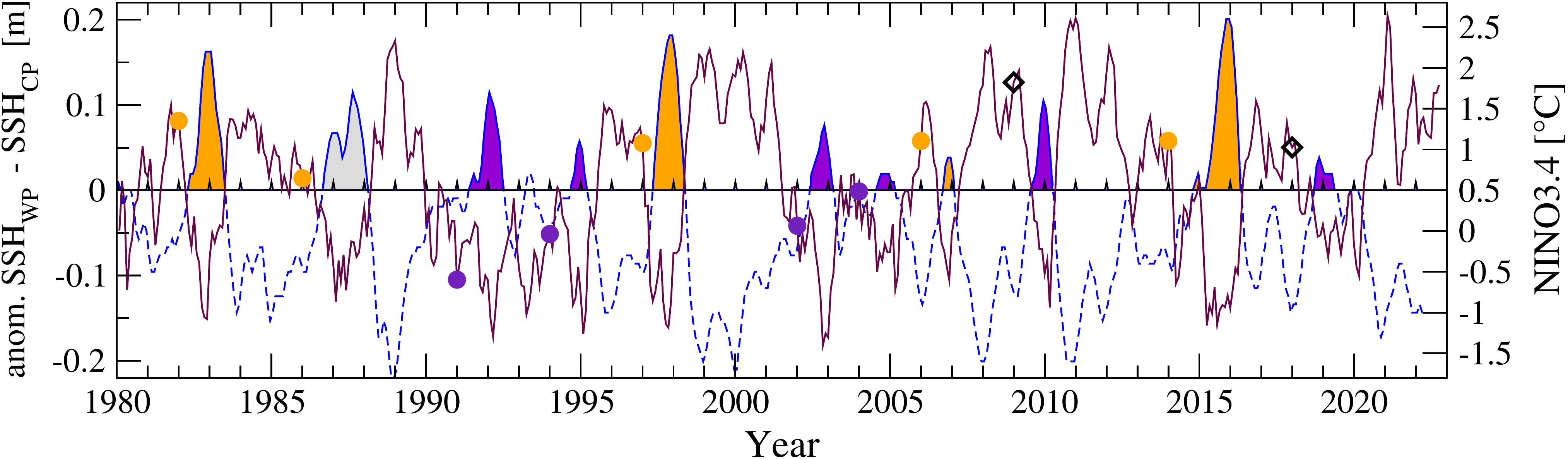}
\caption{SSHA based forecasting scheme for the type of an El~Ni\~no event. Same as Fig. 3 but for the anomaly of the monthly sea surface height (SSH) difference between west Pacific and central Pacific as predictor for the type of an El~Ni\~no. Forecasting based on SSHA leads to the same forecasts as that based on SSTA.}
\label{fig4}
\end{center}
\end{figure*}

Figure 3 shows the difference  $\Delta T_{WP-CP}(t)$ between the mean SSTA  in both areas, as a function of time  $t$ (maroon line), starting in 1950 and ending at present. The blue line is the Oceanic Ni\~no Index (ONI), which is defined as the three-month running-mean SSTA in the Ni\~no3.4 region (see Fig. 2).
The filled areas mark the El~Ni\~no events where the ONI is at least for 5 months greater or equal 0.5°C.
The orange color stands for the 9 EP El~Ni\~no events, the violet color for the 12  CP events, obtained from our consensus classification. For the El~Ni\~no events starting in 1969 and the event starting in 1986 and ending in 1988, there is no consensus, so they are marked gray.

There are two observations on the time evolution of $\Delta T_{WP-CP}(t)$:
(i) $\Delta T_{WP-CP}(t)$ looks like  the mirror image of the ONI, running  through deep minima right at the peak of an El~Ni\~no event and reaching strong maxima in pronounced La~Ni\~na events.
(ii) In the last months of a year preceding an  El~Ni\~no onset, $\Delta T_{WP-CP}(t)$ tends to be positive when the upcoming event is an EP event and negative when it  is a CP event. In the following, we will use this feature as a precursor for the type of an upcoming  El~Ni\~no, with a lead time of about 1 y.

To be specific, we focus on the December values of $\Delta T_{WP-CP}(t)$ for each year between 1950 and 2021. When a December value is positive, we expect that an upcoming El~Ni\~no will be an EP event, otherwise a CP event.
Figure  3 shows that all  9 EP El~Ni\~no events are correctly forecasted by this precursor. There are 3 ``false alarms'',
in 1962,  2008, and 2017, where an EP event is forecasted, but a CP event is observed in the following year.  In contrast, all 9 CP forecasts were correct. Accordingly, when $\Delta T_{WP-CP}(t)$ is negative in December,   we can be highly certain that an upcoming event will be a CP event, i.e., a possibly disastrous EP event in the following year can be excluded with high probability. Otherwise, when $\Delta T_{WP-CP}(t)$ is positive in December, an upcoming El~Ni\~no will be an EP event with 75 percent probability. In October 2022, $\Delta T_{WP-CP}(t)=1.6$. It seems unlikely that $\Delta T_{WP-CP}(t)$ will fall below zero in the next 2 months. Thus, should an El~Ni\~no start in 2023, it probably  will be an EP event.
 
In total, 18 out of 21 events were correctly forecasted. When random guessing with the past occurrence probabilities of EP and CP El~Ni\~no events, the $p$-value for correctly predicting 18 of 21 events is $p=9.1\cdot 10^{-4}$, i.e., the predictions by our precursor are highly significant.

Apart from the zonal SSTA difference, the zonal sea surface height anomaly (SSHA) difference and the zonal surface air temperature anomaly difference are strongly related to the Walker circulation and its fluctuations. Figure 4 shows the difference of the monthly SSHAs between 1980 and present.  The figure shows that the SSHA is highly correlated with the SSTA and leads to the same forecasts. As we show in the SI (SI Fig. 5), the surface air temperature anomalies are also highly correlated with the SSTA and lead to the same forecasting performance.
 
We like to note that our findings are consistent with Ashok et al. \cite{Ashok2007}, where correlation maps for the  El~Ni\~no Modoki Index (EMI) with the SSHA and SSTA in the tropical Pacific have been calculated.
 
\subsection*{Forecasting El~Ni\~no events and their type}
The precursor is most useful when  combined with other methods that forecast the onset of an El~Ni\~no event, irrespective of its type (see, e.g., \cite{Cane86,Penland1993,Palmer2004,Chekroun2011,Saha2014,Chapman2015,
Feng2016,Lu2016,Rodriguez2016,Meng2018,Nootboom2018,Ham2019,Meng2019,DeCastro2020,Petersik2020,Hassanibesheli2022}).
For simplicity, we discuss here  two kinds of forecast:
 
{\bf A. Conventional forecasts.}
Coupled general circulation models (GCM) are initialized by observations and directly simulate the further development of physical quantities like the SSTA. When the predicted development of the ONI, i.e., the SSTA in the Ni\~no3.4 region,  satisfies the definition of an El~Ni\~no, the models predict the onset of an El~Ni\~no event. Since the predictions of these models are limited by the spring barrier, the lead time is about 6 months \cite{Barnston2012, McPhaden2020}.
 
A systematical study  of the performance of the CP/EP prediction  using  GCMs has been done by Ren et al. \cite{Ren2019}, who focused on  6 operational GCMs. All  models were initialized in November. Then the predicted  temporal and spatial distribution of the SSTA in December, January and February is used to forecast whether an El~Ni\~no event will come and whether it will be an EP or CP event.  The result was pretty surprising (see Table 1):
Despite the very short lead time, only about every second EP or CP event was correctly forecasted. The limited skill in distinguishing between EP and CP events  might be due to the models' biases in simulating the background mean state \cite{Guilyardi2012a,Guilyardi2012b,Capotondi2015, Guilyardi2016}.

\begin{figure}[]
\begin{center}
\includegraphics[width=\columnwidth]{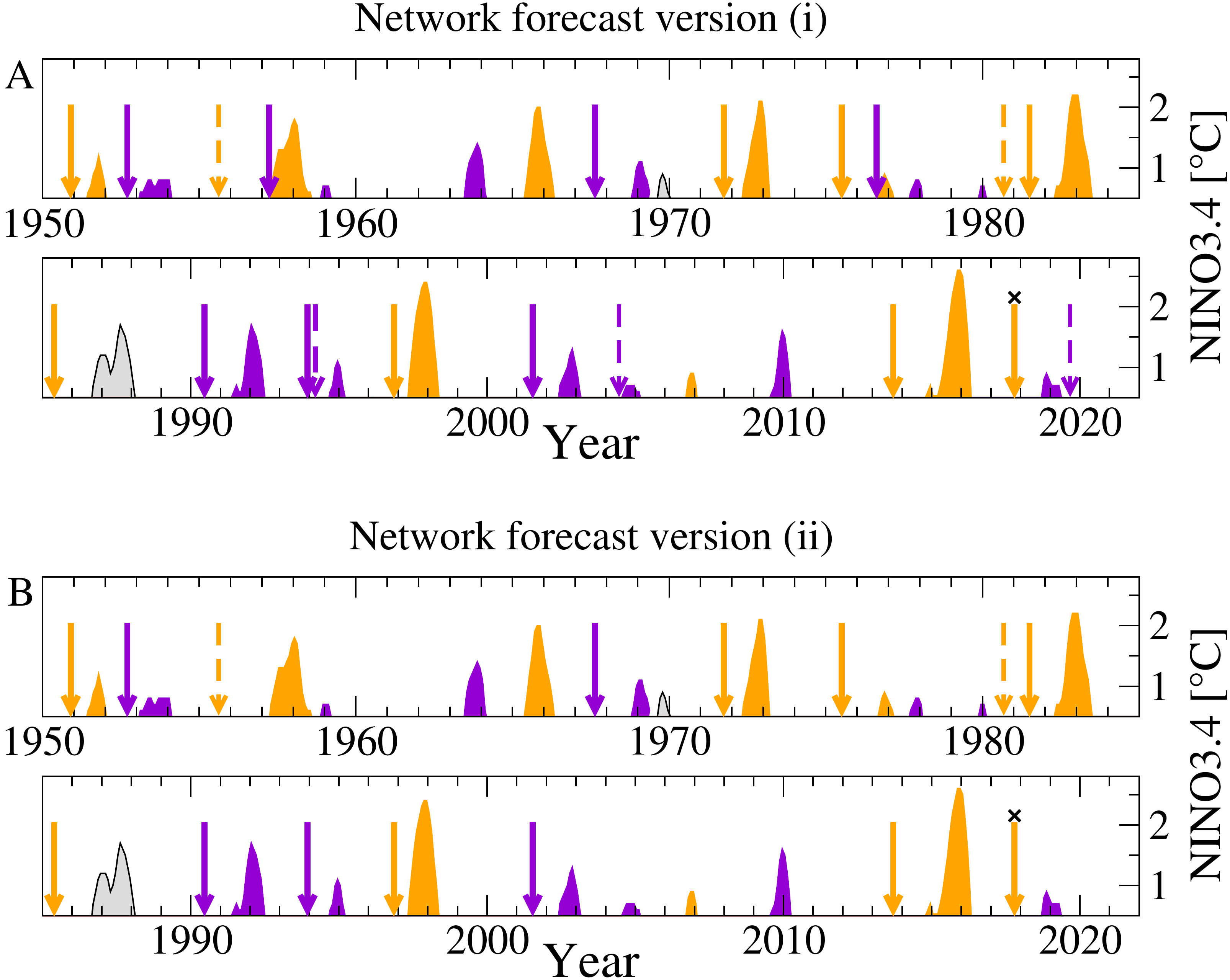}
\caption{Summary of the combined prediction scheme.
Shaded areas mark the El~Ni\~no events: EP (orange), CP (violet), not classified (gray).
Alarms of the network-based method (version (i) in A and version (ii) in B) are combined with the forecast based on the temperature anomaly difference $\Delta T_{WP-CP}(t)$ (Fig. 3)  and shown as arrows in the color of the forecasted El~Ni\~no type.
False alarms of the network-based method are shown as dashed arrows.
The single incorrect forecast of the type (2017) is marked by an 'x'. For better visibility, we have doubled the height of  the 2 smallest CP events, starting in 1958 and 1979.}
\label{fig5}
\end{center}
\end{figure}

{\bf B. Climate network approach.}
The climate network approach \cite{Ludescher2013, Ludescher2014} is described in the Methods Section. As in Fig. 3, we consider the period from 1950 until present. In its original version (i) (see Fig. 5A and SI Fig. 3), the approach yields 20 alarms in the year before the {\it onset} of an El~Ni\~no event; 15 forecasts are correct, including 6 of the 8 largest  El~Ni\~no events  (see SI).  Accordingly, 75\% of the alarms are correct. The total $p$-value of the predictions (including learning, hindcasting, and forecasting phase) calculated from random guessing with the climatological El~Ni\~no onset probability (about 1/3 per year) is $p=3.9\cdot 10^{-7}$.
 
Combining this approach with our precursor shown  in Fig. 3, the method  correctly forecasts 6 EP events (among them the  4 largest ones) and  7 CP events.  When the climate network approach sounded a correct alarm, then also the type of the event was correctly forecasted, with one exception in 2017. Accordingly, when an EP event was forecasted,  the forecast was  about 67\% correct, in 11\% of the cases  a   CP event was observed, and in 22\% of the cases an El~Ni\~no did not occur.
(The correct alarm in 1985 before the non-identified event starting in 1986 was not counted.)  When a CP event was predicted, $70\%$ of the predicted events were CP events and $30\%$ no  El~Ni\~no events.
 
In its more restricted version (ii) (see Fig. 5B and SI Fig. 4), the climate network approach sounds 15 alarms, 13 of which are correct alarms and 2  false alarms.  Compared with version (i), the 2 correct alarms for the smallest El~Ni\~no events  (1958/59 and 1977/78, both CP events) are missing.
Now 86.7\% of the alarms are correct, and for all correct alarms, except that one in 2017, the type of the El~Ni\~no is correctly forecasted. As in version (i), $67\%$ of the predicted EP events were EP events, $11\%$ were   CP events and $22\%$ percent no events. But now, each prediction of a CP event was correct.
 
Table 1 compares the performance of the 2 versions of the climate network approach with the performance of the operational GCMs. It shows that our approach, which is based solely on the zonal SSTA difference between the equatorial western and central Pacific, when combined  with the climate network approach, considerably outperforms the conventional GCM approach, even though its lead time is one order of magnitude longer.
 
\begin{table}
\renewcommand{\arraystretch}{1.2}
\renewcommand\tabcolsep{4pt}
\centering
\begin{center}
\begin{tabular}{|c|c|c|c|c|c|c|}
\hline
\hline
\rule[2.1mm]{0mm}{2.1mm}
Observed type &\multicolumn{2}{c|}{GCM} &\multicolumn{2}{c|}{$\Delta T_{WP-CP}$}
& \multicolumn{2}{c|}{$\Delta T_{WP-CP}$}  \\
&\multicolumn{2}{c|}{} & \multicolumn{2}{c|}{network v(i)} & \multicolumn{2}{c|}{network v(ii)} \\
&\multicolumn{2}{c|}{(1 month)} & \multicolumn{2}{c|}{(11 months)} & \multicolumn{2}{c|}{(11 months)} \\
\hline
& \ EP \ & \ CP \ & \, EP \,  &  CP & \ EP \  & \ CP \  \\
\hline
EP &     49 & 31 & 67  &   0 &67& 0   \\
\hline
CP &     41 & 50 & 11 &  70 &11&  100 \\
\hline
Other &  10 & 19 & 22 &  30 &22 &   0\\
\hline
\end{tabular}
\caption{Contingency table showing the skill of the forecasting methods in predicting the type of an El~Ni\~no event. Tabulated are the percentages of events that are observed to occur, given a method has forecasted an EP or CP El~Ni\~no event. ''Other`` outcomes are neutral or La~Ni\~na events.}
\label{table1a}
\end{center}
\end{table}

\section*{Discussion}
In summary, by studying the  zonal SSTA difference between the equatorial western and central Pacific, we arrived at a precursor which  allows to predict the type of an El~Ni\~no  event  well before the spring barrier with a high accuracy. The  approach does not require fitting procedures, only the sign of the  zonal SSTA difference at the end of a year  matters. When it is positive, an El~Ni\~no event arising  in the following year will probably be an EP event, otherwise a CP event.   Interestingly, the precursor predicted the onset of CP events accurately, while only 75 \% of the EP predictions turned out to be correct.

We may attribute  this difference in the EP/CP predictability to stochastic processes, most likely  westerly wind events (WWE) that typically start in boreal spring and are highly relevant for triggering  El~Ni\~no events.
WWEs can be regarded as state dependent noise, where warmer sea surface temperatures in the western and central Pacific favor more WWEs \cite{Timmermann2018}.
It has been noticed by Jadhav et al. \cite{Jadhav2015} that stronger boreal spring through summer WWEs,
with relatively stronger ocean preconditioning (e.g., large SSHA in the west \cite{Izumo2019}),  can lead to EP events, while weaker ocean preconditioning (small SSHA in the west \cite{Izumo2019}) and weaker WWE can generate CP events. For similar arguments, see, e.g., \cite{Hu2014, Fedorov2015}.

In light of this, we may interpret our results as follows: When the zonal difference of the SSTA (Fig. 3) or the SSHA (Fig. 4) is positive at the end of a year and sufficiently strong WWEs occur in boreal spring in the following year, an  upcoming event will most probably be an EP event, but for weaker WWEs, a CP event may occur with high probability.  This may explain the 25\% incorrect EP forecasts. On the other hand, when the zonal difference is negative, the WWEs can only trigger CP events, and this may explain why we do not see incorrect CP forecasts.
  
For more precise predictions, we have combined the precursor with a climate network approach that forecasts the onset of an  El~Ni\~no in the next calendar year and also does not require a large computational effort.
The climate network  sounds an alarm when its mean link strength exceeds a fixed threshold  and the ONI remains below 0.5 until the end of the year. In this  case, the onset of an  El~Ni\~no event is forecasted for the next calendar year.
From the perspective of predicting the major disastrous events, the combined approach was quite successful. In the last 7 decades, it forecasted the 4 largest  El~Ni\~no events.

Since our El~Ni\~no type prediction approach is not based on a fitting procedure, we are confident that it remains valid also under climate change conditions where the background state of the equatorial Pacific might change.
We hope that  the approach will enable early  and more targeted mitigation methods, either together with the climate network approach discussed here or with any other early forecasting approach, and thus help prevent or at least  mitigate humanitarian disasters as consequences of El~Ni\~no related extreme weather impacts.
   
\section*{Data and Methods}

\noindent{\bf Data.}
The monthly sea surface temperatures were obtained from the National Oceanic and Atmospheric Administration (NOAA) Extended Reconstructed Sea Surface Temperature version 5 (ERSSTv5) \cite{ERSST5}.
The daily surface air temperature (SAT) data (1948-present) for the calculation of the mean link strength $S(t)$ of the network (see below) were obtained from the National Centers for Environmental Prediction/National Center for Atmospheric Research (NCEP/NCAR) Reanalysis I project \cite{reanalyis1,reanalyis2}. For the sea surface height (SSH), we used the monthly NCEP Global Ocean Data Assimilation System (GODAS) \cite{Behringer1998} data set obtained from \cite{ERDDAP}.
\bigskip\noindent

\noindent{\bf Climate Network Approach.}
The  approach exploits the  observation that a large-scale cooperative mode linking the ``El~Ni\~no basin'' (i.e., the equatorial Pacific corridor) and the rest of the tropical Pacific (see SI Fig. 2 and \cite{Gozolchiani2011,Ludescher2013})
builds up in the calendar year before an El~Ni\~no event.   The emerging cooperativity is derived from the time evolution of the teleconnections (``links``) between the SATA at the grid points (''nodes``) inside and outside of the El~Ni\~no basin. The strengths of these links at a given time $t$ are derived from the values of the respective cross-correlations (for details, see \cite{Gozolchiani2011,Ludescher2013}). The  mean link strength $S(t)$ in the network usually  rises  in the year before an El~Ni\~no event starts  and drops  with the onset of the event. This feature serves as a precursor for the event.

The algorithm \cite{Ludescher2013} involves as only fit parameter a decision threshold $\Theta$, which  has been fixed in a learning phase between 1950 and 1980.   In its  original version (i), the algorithm gives an alarm and predicts an El~Ni\~no inception in the following year whenever $S$ crosses $\Theta$ from below while the most recent ONI is below 0.5°C.
In a modified version (ii), the algorithm considers only those alarms for which the ONI remains below 0.5 for the rest of the calendar year.

\bigskip\noindent
{\bf Acknowledgments} J.L. thanks the “Brazil East Africa Peru India Climate Capacities (B-EPICC)” project, which is part of  the International Climate Initiative (IKI) of the German Federal Ministry for Economic Affairs and Climate Action (BMWK) and implemented by the Federal Foreign Office (AA).

\end{document}


\maketitle

\section{El~Ni\~no type classification}

For the classification of the type of an El~Ni\~no, we consider 11 methods, which were referred to in previous classification summaries \cite{Yu2012, Wiedermann2016, Zheng2017, Feng2019, Capotondi2020}, and show them in SI Fig. 1. For the methods of Larkin et al. 2005 \cite{Larkin2005}, Feng et al. 2019 \cite{Feng2019}, Wiedermann et al. 2016 \cite{Wiedermann2016}, Yu and Kim 2013 \cite{YuKim2013}, Kim 2009 \cite{Kim2009}, and Yeh et al. 2009 \cite{Yeh2009} we use the original publication results. Some of the original studies did not provide a classification list or were considerably extended in later studies. Thus for the classification according to the El~Ni\~no Modoki index (EMI; Ashok et al. 2007 \cite{Ashok2007}), we show the results of \cite{Zheng2017} and \cite{Feng2019}. For the EP-CP index method (Kao \& Yu 2009, \cite{KaoYu2009}), we show the results from the table in \cite{Yu2012} and fill up the remaining years from the website of Yu \cite{websiteYu}.
For the cold tongue (CT) and warm pool (WP) index method of Ren and Jin (2011) \cite{RenJin2011}, we use the table in \cite{Zheng2017}.
For the E and C indices (Takahashi et al. 2011 \cite{Takahashi2011}), we use \cite{Capotondi2020}. Here, years where the classification changes between November-December-January and December-January-February, are marked ''M'' (yellow). For Kug et al. 2009 \cite{Kug2009}, we extended the original publication by the table in \cite{Feng2019}. For the Wiedermann et al. method, we filled up 2014 and 2015 via the newer publication \cite{Wiedermann2021}.

The small 2018/2019 El~Ni\~no has not yet been included in classification lists. Here we calculate based on the Ni\~no1+2, Ni\~no3 and Ni\~no4 indices from \cite{NOAA_ERSST} and \cite{NOAA_HadSST} classifications for it. The respective ERSSTv5 NDJ values are +0.90°C, +0.81°C, +0.86°C. Since Ni\~no4 $>$ 0.5°C and Ni\~no4 $>$ Ni\~no3, according to the method of Yeh et al. 2009 the event is a CP one. The Ni\~no1+2, Ni\~no3 and Ni\~no4 values from HadSST \cite{NOAA_HadSST} are +0.72, +0.86°C and +0.72°C, respectively. Since Ni\~no3 $>$ Ni\~no4, this dataset indicates an EP El~Ni\~no. Analogously, the classification via the Kug et al. 2009 method, which also compares the values of Ni\~no3 and Ni\~no4, provides, in this case, opposite results depending on the dataset. Thus we mark in SI Fig. 1 the year 2018 for both methods as not pure EP or CP, i.e., ``M'' (yellow).
The approximate equations for the C and E index (Takahashi et al. 2011, \cite{Takahashi2011}) are: $C\approx 1.7 Nino_4 - 0.1 Nino_{1+2}$ and $E\approx Nino_{1+2} - 0.5 Nino_4$. This yields 1.37 and 0.47 for the C and E index, respectively, signaling a CP El~Ni\~no. For the HadSST data E=0.36 and C=1.17, confirming a CP event.

For the Kao and Yu 2009 method, we obtained additional values from the website of JY Yu \cite{websiteYu}. We consider the December values of the indices: 1979: CP= 1.118, EP= 0.326 $\Rightarrow$ CP-event, 2014: CP=0.713, EP=-0.421 $\Rightarrow$ CP-event, 2015: CP= 1.679 EP=1.997 $\Rightarrow$ EP-event, 2018: CP=0.443, EP=-0.184 $\Rightarrow$ CP-event.

We consider the 1986-1988 and the 2014-2016 2 year El~Ni\~no events as single continuous events, respectively. Several studies considered and classified them as two separate events, which is reflected in the shown onset years in the table (SI Fig. 1). However, our consensus is based on the average classification over the full 2 y period.

\begin{figure}[]
\begin{center}
\includegraphics[width=12cm]{./fig1.pdf}
\caption{Classification of the type of El~Ni\~no events. The table summarizes the classification according to 11 approaches. The numbers in the first row indicate the year of the onset of an El~Ni\~no.
Eastern Pacific El~Ni\~nos are marked 'E' and shown in orange. Central Pacific El~Ni\~nos are marked 'C' and shown in blue. Classifications shown in light yellow and marked 'M' are considered by the respective method as: mixed events, events where the type is not consistent over the course of the event, and events where the classification is not consistent over different data sets, see main Supporting Information text.}
\label{figS1}
\end{center}
\end{figure}

\section{Network-based El~Ni\~no onset prediction}

For the description of the climate network-based forecasting algorithm \cite{Ludescher2013} for the onset of El~Ni\~no events, we follow \cite{Ludescher2014}. The algorithm is as follows:

(1) At each node $k$ of the network shown in SI Fig. 2 the daily atmospheric temperature anomalies $T_k(t)$ (actual temperature
value minus climatological average  for each calendar day, see below) at the surface area level are determined. For simplicity, leap days are removed. The data were obtained from the National Centers for Environmental Prediction/National Center for Atmospheric Research Reanalysis I \cite{reanalyis1,reanalyis2}.

(2) For obtaining the time evolution of the strengths of the links between the nodes $i$ inside the El~Ni\~no basin and the nodes $j$
outside, we compute, for each 10th day $t$ in the considered time span between January 1950 and present, the time-delayed cross-correlation
function defined as
\begin{equation}
 C_{i,j}^{(t)}(-\tau)=\frac{\langle T_i(t)T_j(t-\tau)\rangle-\langle T_i(t)\rangle\langle T_j(t-\tau)\rangle}{\sqrt{\langle(T_i(t)-\langle T_i(t)\rangle)^2 \rangle}\cdot\sqrt{\langle(T_j(t-\tau)-\langle T_j(t-\tau)\rangle)^2 \rangle}}
\end{equation}
and
\begin{equation}
 C_{i,j}^{(t)}(\tau)=\frac{\langle T_i(t-\tau)T_j(t)\rangle-\langle T_i(t-\tau)\rangle\langle T_j(t)\rangle}{\sqrt{\langle(T_i(t-\tau)-\langle T_i(t-\tau)\rangle)^2 \rangle}\cdot\sqrt{\langle(T_j(t)-\langle T_j(t)\rangle)^2 \rangle}}
\end{equation}
where the brackets denote an average over the past 365 d, according to
\begin{equation}
 \langle f(t) \rangle = \frac{1}{365} \sum_{m=0}^{364} f(t-m).
\end{equation}
We consider time lags $\tau$ between 0 and
200 d, where a reliable estimate of the background noise level can
be guaranteed.

(3) We determine, for each point in time $t$, the maximum, the mean, and the standard deviation around the mean of the absolute value of the cross-correlation function
 $|C_{ij}^{(t)}(\tau)|$
 and define the link
strength $S_{ij}(t)$ as the difference between the maximum and the
mean value, divided by the standard deviation. Accordingly, $S_{ij}$ describes the
link strength at day t relative to the underlying background noise (signal-to-noise ratio) and
thus quantifies the dynamical teleconnections between nodes i and j.

(4) To obtain the desired mean strength $S(t)$ of the dynamical teleconnections in the climate network, we simply average over all individual link strengths.

For the calculation of the climatological average in the learning phase (1950-1980), all data within this time window are taken into account, whereas in the prediction (i.e., hindcasting and forecasting) phase, only data from the past up to the prediction date are considered.

\begin{figure}[]
\begin{center}
\includegraphics[width=8cm]{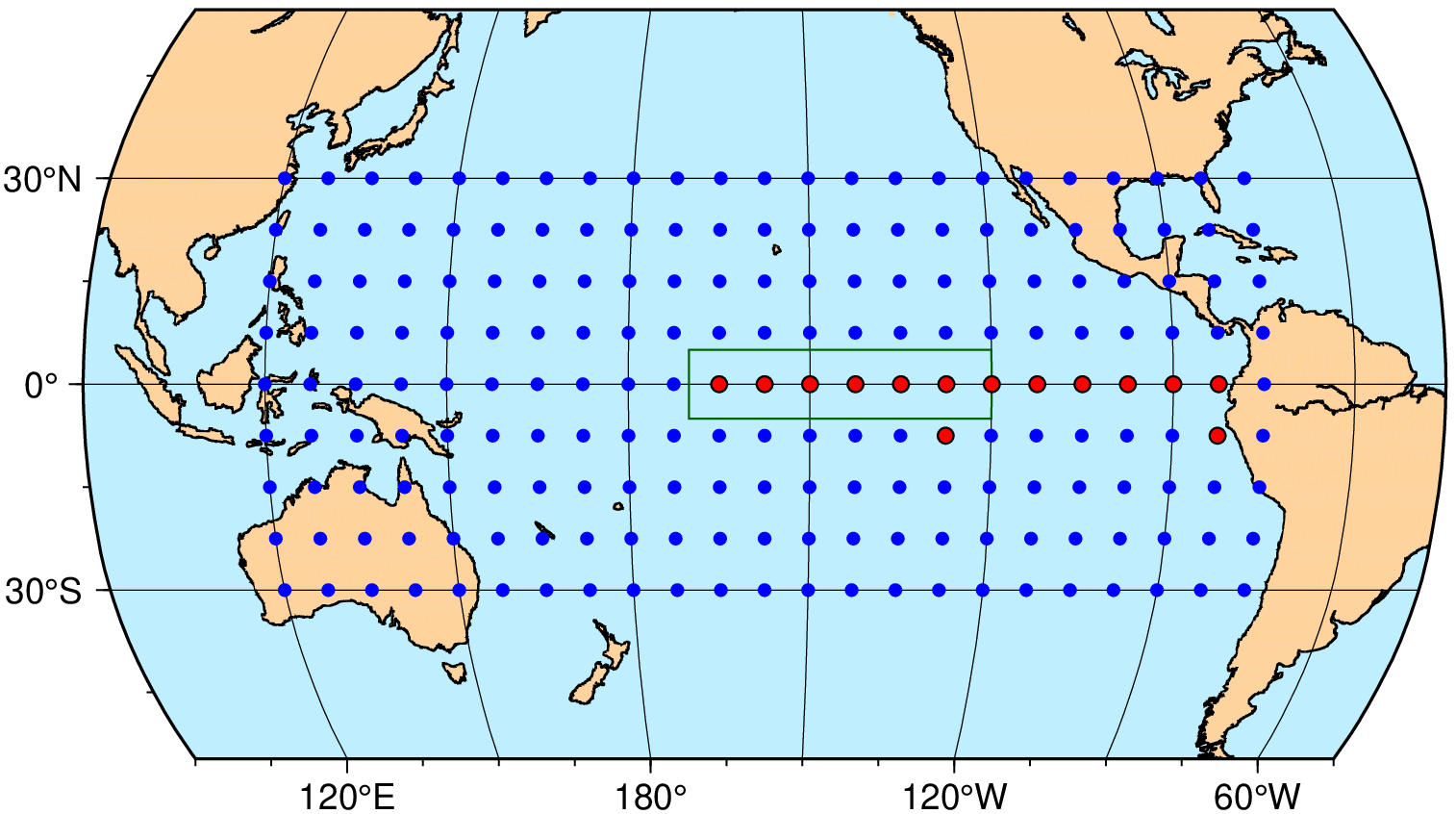}
\caption{The  ONI and the climate network. The network consists of 14 grid
points  in the ``El~Ni\~no basin'' (red dots) and 193 grid points outside this
domain (blue dots). The green rectangle denotes the area where the ONI (Ni\~no3.4 index) is measured. The grid points are considered as the nodes of the climate network that we use here to forecast the onset of El~Ni\~no events. Each node inside the El~Ni\~no basin is linked to each node outside the basin. The nodes are
characterized by their surface air temperature (SAT), and the link strength between the nodes is determined from their cross-correlation (see below).}
\label{figS2}
\end{center}
\end{figure}

\begin{figure}[h!]
\begin{center}
\includegraphics[width=12cm]{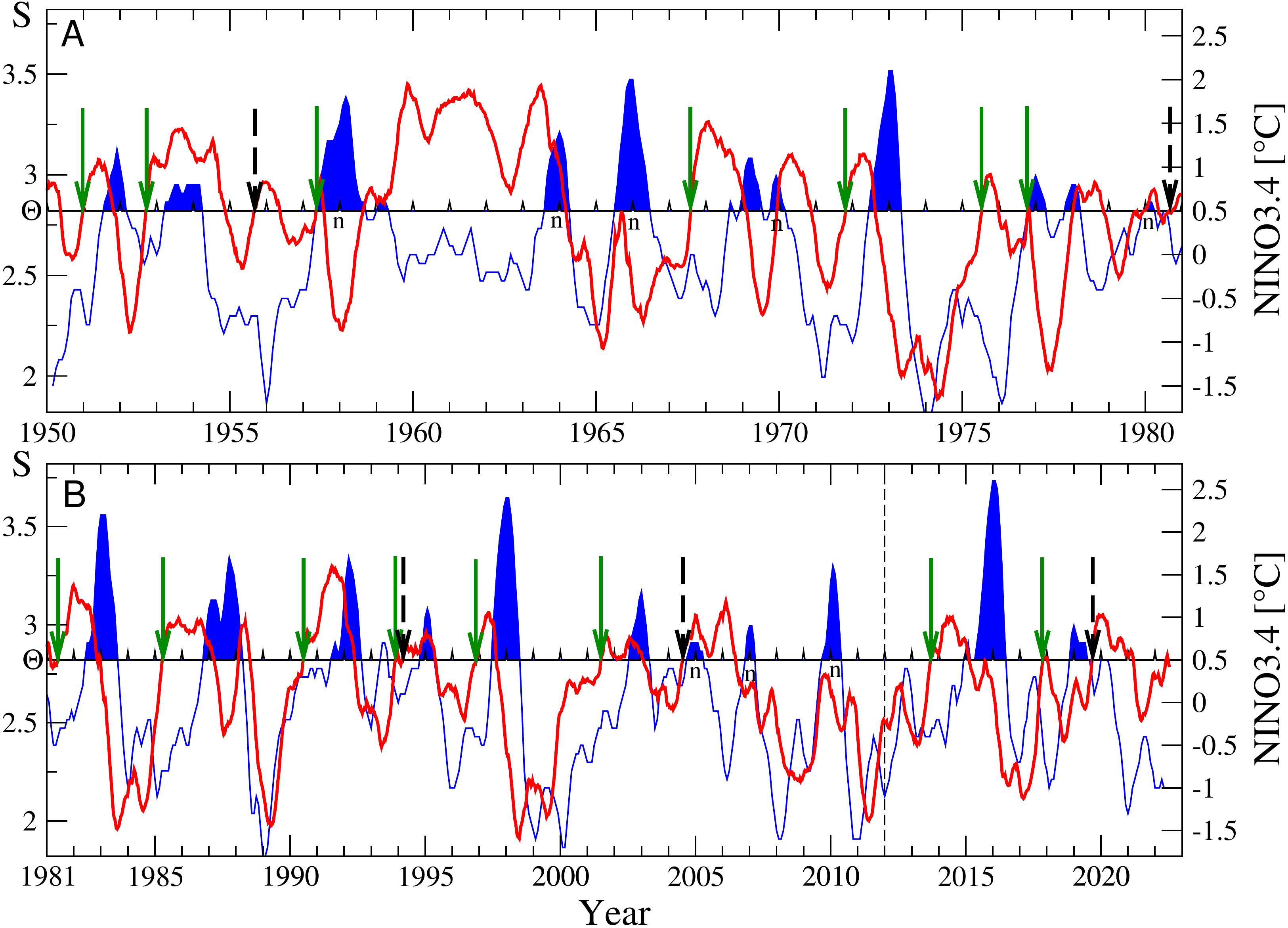}
\caption{
The network-based forecasting scheme version (i). We compare the average link strength $S(t)$
in the climate network (red curve) with a decision threshold $\Theta$ (horizontal line, here $\Theta = 2.82$), (left scale), and the standard Ni\~no3.4 index (ONI), (right scale).
When the link strength crosses the threshold from below, and the last available ONI is below 0.5°C,
we give an alarm and predict that an El~Ni\~no episode will start in the following calendar year.
The El~Ni\~no episodes (when the Ni\~no3.4 index is at or above 0.5°C for at least 5 months) are shown by the solid blue areas.
Correct predictions are marked by green arrows and false alarms by dashed black arrows.
The optimal decision threshold $\Theta$ was determined in a learning phase (1950-1980).
Between 1981 and 2021 (hindcasting and forecasting phases, separated by a dashed vertical line), there were 11 El Ni\~no events. The algorithm generated 11 alarms, and 8 were correct.
In the whole period between 1950 and 2021, there were 23 El Ni\~no events. The algorithm generated 20 alarms and 15 of these were correct.
}
\label{figS3}
\end{center}
\end{figure}

\begin{figure}[h!]
\begin{center}
\includegraphics[width=12cm]{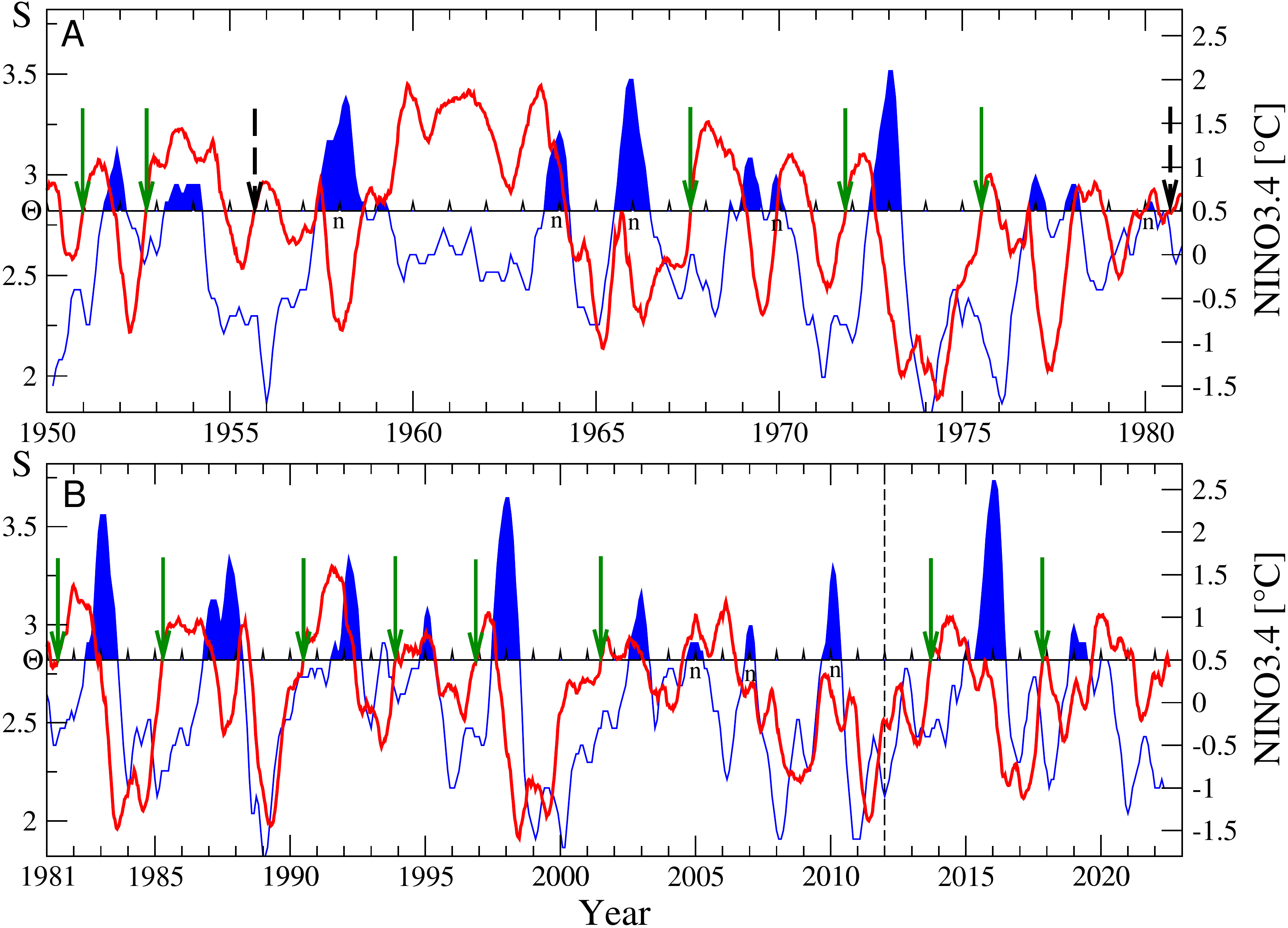}
\caption{
Same as SI Fig. 3, but for version (ii) of the network-based forecasting scheme. Here, only alarms where the ONI remains below 0.5°C until the end of the year are considered. Alarms of version (i) that are not present in version (ii) are shown in gray.
}
\label{figS4}
\end{center}
\end{figure}

\begin{figure}[h!]
\begin{center}
\includegraphics[width=12cm]{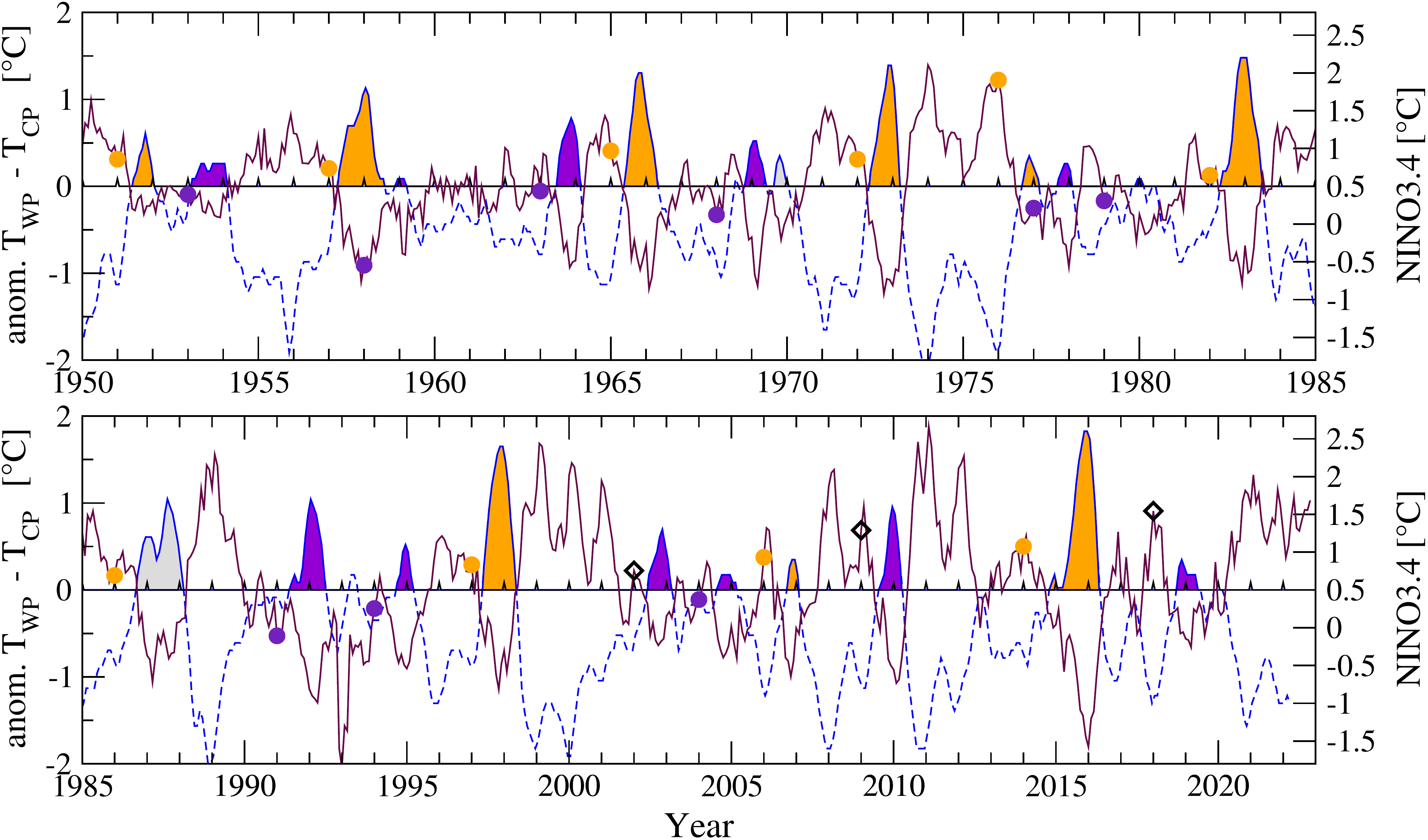}
\caption{
SATA-based forecasting scheme for the type of an El~Ni\~no event. Same as Fig. 3 in the main text but for the monthly surface air temperature anomalies (SATA) obtained from
the National Centers for Environmental Prediction/National Center for Atmospheric Research (NCEP/NCAR) Reanalysis I project \cite{reanalyis1,reanalyis2}.
Forecasting based on SATA leads to the same forecasts as based on SSTA or SSHA except in 1962, where now a CP El~Ni\~no is correctly forecasted, and 2001, where now an EP El~Ni\~no is incorrectly forecasted.
}
\label{figS5}
\end{center}
\end{figure}